\documentclass{article}

\PassOptionsToPackage{numbers, compress}{natbib}

\usepackage[preprint]{SMARTFinRAG}

\usepackage[utf8]{inputenc} 
\usepackage[T1]{fontenc}    
\usepackage{hyperref}       
\usepackage{url}            
\usepackage{booktabs}       
\usepackage{amsfonts}       
\usepackage{nicefrac}       
\usepackage{microtype}      
\usepackage{xcolor}         
\usepackage{graphicx}
\usepackage{listings}
\usepackage{xcolor}
\usepackage{tabularx}  
\usepackage{makecell}  
\usepackage{pifont}
\newcommand{\cmark}{\ding{51}} 
\newcommand{\xmark}{\ding{55}} 
\usepackage{tabularx}
\usepackage{makecell}
\usepackage{amsmath}

\title{SMARTFinRAG: Interactive Modularized Financial RAG Live-Demo System}

\author{
 Yiwei Zha \\
  Khoury College of Computer Science\\
  Northeastern University\\
  Boston, MA \\
  \texttt{zha.y@northeastern.edu} \\
}

\begin{document}

\maketitle

\begin{abstract}
Financial sectors are rapidly adopting language model technologies, yet the evaluation of specialized Retrieval-Augmented Generation (RAG) systems in this domain remains challenging. Financial applications demand exceptional accuracy, contextual relevance, and adaptability to rapidly changing information—requirements not adequately addressed by current evaluation approaches.\\
This paper introduces SMARTFinRAG, a comprehensive framework designed to address three critical gaps in financial RAG assessment. First, we implement a fully modular architecture where components can be dynamically interchanged during runtime, allowing researchers to experiment with various retrieval mechanisms (BM25, vector-based, hybrid), language models, and prompt strategies. Second, we develop a document-centric evaluation paradigm that generates domain-specific question-answer pairs from newly ingested financial documents, enabling continuous assessment with current information. Third, we bridge research-implementation divides through an intuitive interface where practitioners can visualize performance metrics and observe real-time system behavior.\\
Our evaluation demonstrates how SMARTFinRAG effectively quantifies both retrieval efficacy (through hit rate, MRR, NDCG metrics and etc.) and response quality (via LLM-as-judge faithfulness/relevance assessment), revealing significant performance variations across configurations. The platform's open-source architecture\footnote{\url{https://github.com/JonathanZha47/SMARTFinRAG}} supports transparent, reproducible research while addressing practical deployment challenges faced by financial institutions implementing RAG systems.
\end{abstract}

\section{Introduction}
The financial sector has witnessed substantial transformation through the application of large language models (LLMs), which now power sophisticated natural language processing capabilities across question answering, earnings report analysis, and market sentiment evaluation. Within this evolving landscape, Retrieval-Augmented Generation (RAG) has emerged as a critical methodology that enhances LLM outputs by anchoring responses in verifiable external sources. This evidence-grounding approach holds particular significance in financial contexts, where information precision, currency, and source verification directly impact decision-making processes.

Despite significant advancements in financial NLP applications, robust evaluation frameworks for financial RAG systems remain severely limited. Current approaches fall into two principal categories: financial question-answering datasets like FiQA~\cite{fiqa2018} and FinQA~\cite{chen2021finqa} that lack retrieval integration capabilities, or comprehensive financial benchmarks such as FinBen~\cite{xie2024finben} and PIXIU~\cite{xie2023pixiu} that primarily focus on static model capabilities rather than end-to-end RAG pipeline assessment. Meanwhile, general-purpose RAG evaluation frameworks like RAGAS~\cite{shahul2023ragas} and ContextCheck~\cite{contextcheck2024} offer valuable metrics but require substantial adaptation for finance-specific requirements including numerical accuracy, temporal sensitivity, and regulatory compliance. Neither approach sufficiently addresses the need for dynamic component evaluation, real-time assessment with updated financial information, or interactive experimentation capabilities crucial for bridging the gap between research prototypes and production-ready financial systems.

To address these limitations, we present \textbf{SMARTFinRAG}, an interactive evaluation platform specifically engineered for financial RAG assessment. Our framework delivers three key innovations that differentiate it from existing solutions:

First, SMARTFinRAG implements a comprehensive parameter configuration system that enables dynamic adjustment of both retrieval components (retriever selection, context window size, ranking weights) and generation parameters (model selection, temperature, prompt strategies). This capability facilitates multidimensional experimental evaluation without requiring system reconfiguration.

Second, the platform embraces a fully modularized architecture where each RAG pipeline component functions as an independently replaceable unit. This design allows researchers and practitioners to selectively activate or deactivate specific elements, compare alternative implementations, and identify performance bottlenecks through controlled experimentation.

Third, SMARTFinRAG introduces a document-centric evaluation paradigm that leverages an LLM-as-Judge approach to generate domain-appropriate question-answer pairs directly from ingested financial documents. This methodology enables continuous assessment using the most current information, with comprehensive metrics capturing both retrieval quality (precision, recall, MRR) and response fidelity (faithfulness, relevance) within a unified evaluation framework.

Beyond serving as a benchmarking tool, SMARTFinRAG functions as a reusable \emph{RAG demonstration paradigm} for financial question-answering research. Its interactive web interface supports rapid prototyping, component ablation studies, and reproducible evaluation—all grounded in authentic financial document processing. Through this approach, we aim to accelerate both academic research and practical implementation of trustworthy financial RAG systems.

\textbf{Our key contributions include:}
\begin{itemize}
    \item Development of a pioneering financial domain-specific RAG evaluation platform that accommodates both historical document analysis and real-time information integration, addressing unique challenges in financial information processing.
    
    \item Introduction of an innovative assessment methodology centered on document-derived benchmarks, which simultaneously evaluates information retrieval accuracy and response generation quality through complementary metrics targeting lexical precision, semantic relevance, and fabrication detection.
    
    \item Creation of an intuitive browser-based demonstration environment enabling immediate configuration adjustments, component substitution, performance visualization, and exportable evaluation reports for research reproducibility.
    
    \item Implementation of a standardized question-answering framework that integrates diverse financial data sources while providing granular diagnostic capabilities to precisely locate performance limitations within complex RAG pipelines.
\end{itemize}

Collectively, these advancements establish a practical research and evaluation platform that accelerates the development of reliable, evidence-grounded language models for financial applications, while supporting systematic experimentation at the convergence of information retrieval and natural language generation technologies.

In the following sections, the section 2 will talk about the related works, section 3 introduces the detailed methodology of the framework, and section 4 will analyze the experimental results.

\section{Related Work}

\subsection{Financial RAG Systems.}
Recent advancements have introduced Retrieval-Augmented Generation (RAG) systems tailored for the financial domain. For instance, FinQAPT~\cite{singh2024finqapt} presents an end-to-end pipeline that streamlines the identification of relevant financial reports based on a query, extracts pertinent context, and leverages Large Language Models (LLMs) to perform downstream tasks. However, its evaluation pipeline is limited to closed QA settings and lacks support for real-time document ingestion, dynamic retriever switching, or detailed metric instrumentation. Similarly, FinArena~\cite{xu2025finarena} employs an adaptive RAG method for processing unstructured news data, aiming to address hallucinations in LLMs. Yet, its architecture is task-specific and not designed for modular benchmarking across LLMs or retrievers. Both systems offer limited evaluator flexibility and do not expose configurable pipelines for systematic experimentation. They also lack user-facing interfaces for interactive testing or diagnostic visualization. In contrast, SMARTFinRAG provides a live demo interface that supports real-time document ingestion, plug-and-play retriever selection, and multi-metric evaluation, offering a more extensible and researcher-oriented platform.

\subsection{Challenges in General RAG Systems.}
Retrieval-Augmented Generation (RAG) systems have demonstrated potential in enhancing the factual grounding of large language models (LLMs) by conditioning generation on retrieved evidence. However, several systemic challenges persist. These include persistent hallucinations, particularly when retrieval returns incomplete or noisy passages; fragile pipeline behavior when faced with retrieval uncertainty; and lack of calibration or interpretability in retriever scores.

As highlighted by \cite{church2024emerging}, RAG systems face fundamental limitations in addressing several critical issues. Their research demonstrates that LLMs struggle with ``hallucinations when asked to discuss content that goes beyond training data,'' which RAG aims to mitigate but cannot fully eliminate, especially when guard rails are easily circumvented by slight prompt modifications. The timeliness problem remains particularly challenging, as ``it is prohibitively expensive for [providers] to continuously update their models,'' making just-in-time knowledge updating crucial but imperfect in implementation.

Furthermore, \citeauthor{church2024emerging} identify document processing limitations where ``RAG tends to process documents in very simple ways, typically as a sequence of chunks,'' failing to properly handle complex document structures containing tables, figures, and other elements. This simplistic approach, combined with OCR errors in older documents, significantly impacts retrieval quality. Their evaluation also reveals that RAG summaries, while impressive at first glance, are often ``long-winded'' and ``sometimes seem to miss the point,'' indicating fundamental limitations in how these systems integrate retrieved information.

While open-source frameworks such as LangChain and LlamaIndex have introduced modules for evaluating retrieval quality and response faithfulness, these tools often require additional configuration and may not fully address the stringent requirements of high-stakes domains like finance. Moreover, few systems support real-time indexing or dynamic corpus updates, limiting their effectiveness in fast-changing environments. SMARTFinRAG addresses these issues by providing an interactive platform that allows users to experiment with different RAG components in real-time, facilitating a deeper understanding of system behavior.

\subsection{Limitations of Financial LLM Benchmarks.}
While several benchmarks have been developed to evaluate the capabilities of large language models (LLMs) in the financial domain, significant limitations persist. For instance, FinBen~\cite{xie2024finben} encompasses 35 datasets across 23 tasks, aiming to assess a broad spectrum of financial competencies. Similarly, FinanceBench~\cite{islam2023financebench} offers a comprehensive suite for financial question answering but has been criticized for its limited coverage of complex reasoning tasks and potential data leakage issues. Furthermore, benchmarks like FailSafeQA~\cite{names2025failsafeqa} have highlighted the prevalence of hallucinations in financial LLM outputs, noting that models can produce incorrect information in up to 41\% of finance-related queries. These shortcomings underscore the need for more robust, dynamic, and comprehensive evaluation frameworks that can accurately assess LLM performance in high-stakes financial applications.

To make these differences more concrete, Table~\ref{tab:rag-comparison} provides a comparison of existing financial NLP tools and frameworks across four core capabilities: financial data support, modularity, real-time evaluation, and demo-readiness.

\begin{table}[ht]
\centering
\small
\renewcommand{\arraystretch}{1.1}
\begin{tabularx}{\linewidth}{lXXXX}
\toprule
\textbf{Tool} & \makecell{\textbf{Financial}\\\textbf{Data Support}} & \makecell{\textbf{Modular}\\\textbf{RAG}} & \makecell{\textbf{JIT}\\\textbf{Evaluation}} & \makecell{\textbf{Demo}\\\textbf{Ready}} \\
\midrule
FiQA / Financial NLP Datasets & \cmark & \xmark & \xmark & \xmark \\
LangChainHub / RAGAS & \xmark & \cmark & \xmark & \xmark \\
\textbf{SMARTFinRAG (Ours)} & \cmark & \cmark & \cmark & \cmark \\
\bottomrule
\end{tabularx}
\caption{Comparison of related tools across core RAG evaluation dimensions. SMARTFinRAG is the only framework supporting modularity, live evaluation, and demo-ready interaction within financial QA workflows.}
\label{tab:rag-comparison}
\end{table}

\section{Methodology}

\subsection{System Architecture}
\begin{figure}[ht]
\centering
\includegraphics[width=0.65\linewidth]{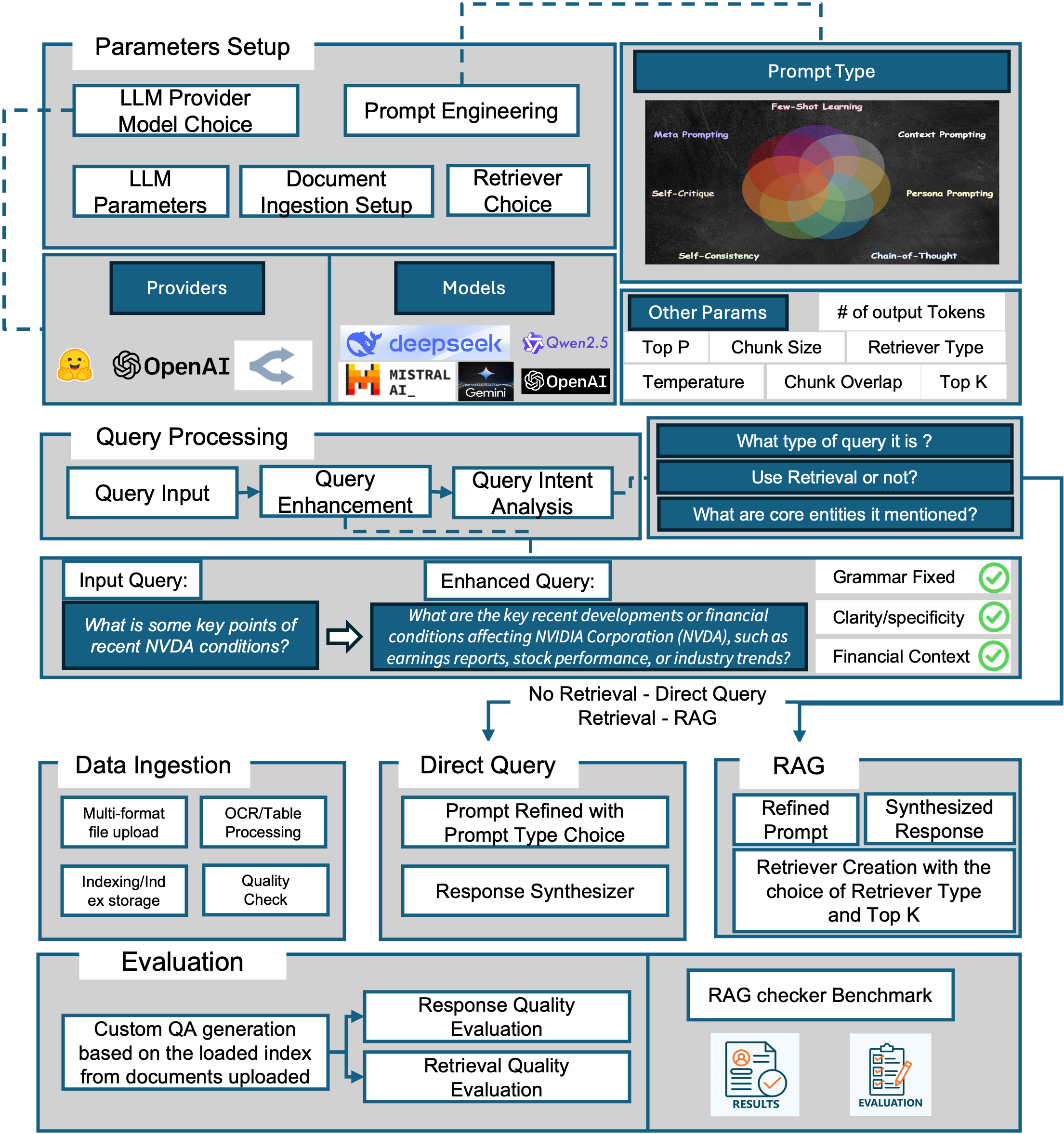}
\caption{SMARTFinRAG pipeline. Modular components process raw financial documents into indexed chunks and support retrieval-augmented response generation with multi-faceted evaluation. Each stage is configurable and replaceable.}
\label{fig:architecture}
\end{figure}

SMARTFinRAG is a modular and extensible framework designed to benchmark financial Retrieval-Augmented Generation (RAG) systems under realistic conditions. The overall architecture consists of independently configurable components organized into a composable pipeline. As shown in Figure~\ref{fig:architecture}, the system supports end-to-end processing from document ingestion to response evaluation.

\textbf{Pipeline Overview.} The architecture is composed of five primary stages:
\begin{enumerate}
    \item \textbf{Document Ingestion}: Parses structured (e.g., TXT, DOCX) and unstructured (e.g., scanned PDF) documents, extracts text and tables, applies OCR when necessary, and segments content into overlapping chunks. These are embedded using OpenAI's embedding model and stored in a persistent vector index.
    
    \item \textbf{Query Preprocessing and Routing}: Enhances the user's question using DeepSeek, an instruction-following LLM selected for its superior analogy and inference capabilities. The system analyzes query intent to determine whether document retrieval is required.
    
    \item \textbf{Document Retrieval}: Implements multiple retrieval strategies including BM25, vector-based (FAISS), hybrid fusion with dynamic weight assignment between BM25 and vector scores, and auto-merging for hierarchical document structures. The appropriate retriever is selected using a factory pattern to fetch the top-$k$ relevant chunks from the indexed corpus.
    
    \item \textbf{LLM-Based Generation}: Passes the enhanced query and retrieved context to a configurable LLM module to synthesize the final answer. Supported models include GPT-3.5 Turbo, GPT-4o-mini, GPT-4o, DeepSeek, Mixtral 7×8B, Google Gemini-Flash-2.0, Gemma3, Qwen-32B, and Llama4-Scout, with specialized templated prompts and grounding instructions.
    
    \item \textbf{Evaluation Engine}: Scores the output using multiple criteria including retriever performance (MRR, NDCG, etc.), LLM-based faithfulness and relevancy, and logs all traces for auditability and visualization.
\end{enumerate}

\begin{figure}[ht]
\centering
\includegraphics[width=0.65\linewidth]{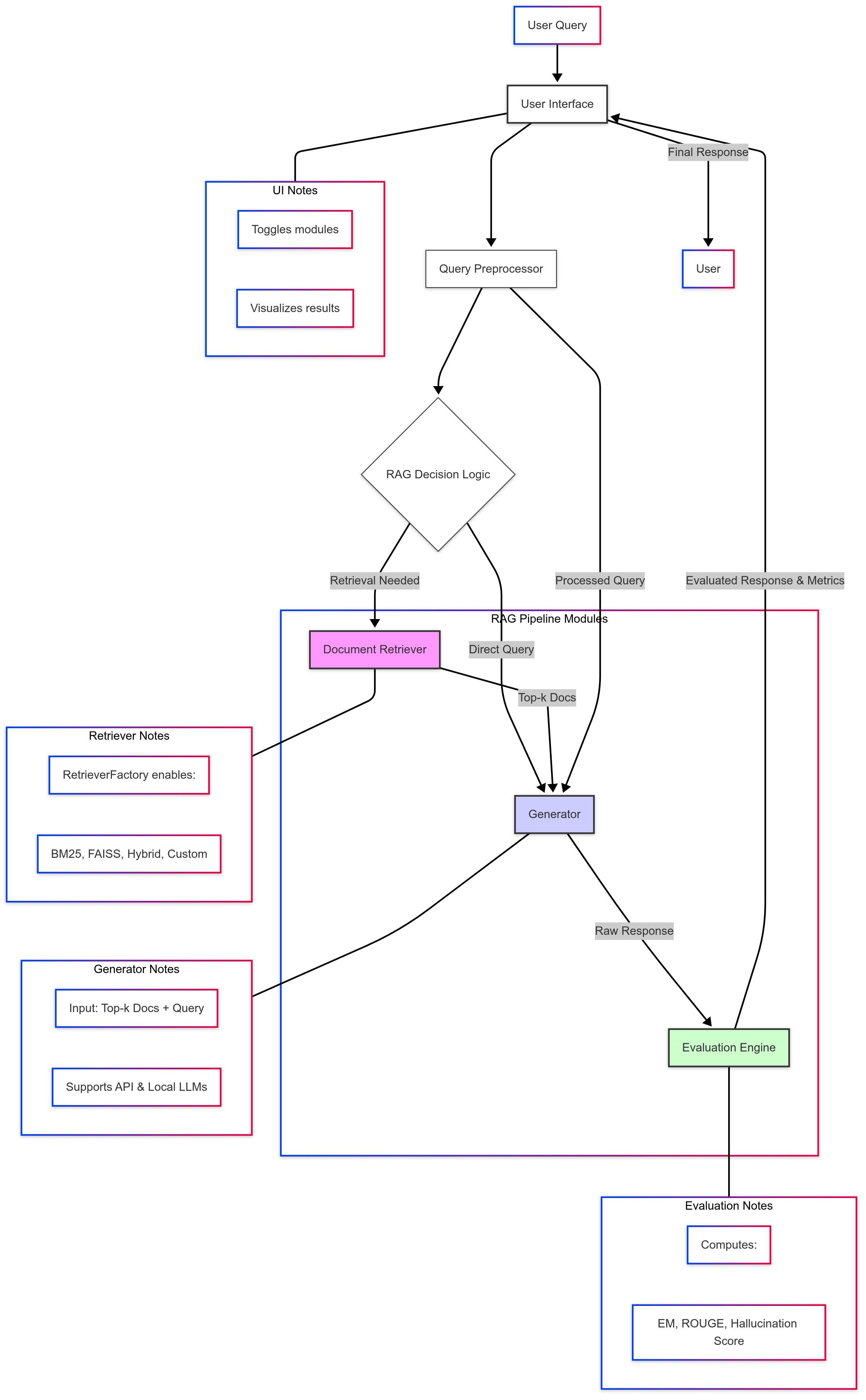}
\caption{SMARTFinRAG pipeline. Modular components process raw financial documents into indexed chunks and support retrieval-augmented response generation with multi-faceted evaluation. Each stage is configurable and replaceable.}
\label{fig:flowchart}
\end{figure}

\textbf{Execution Flow.} As shown in Figure~\ref{fig:flowchart}, a user query is submitted via the web UI or programmatic API. The query is rewritten, classified, and conditionally routed to the retriever. Retrieved passages are forwarded to the generator alongside the rewritten query. The generator produces a grounded response, which is then passed to the evaluator for scoring. All modules can be replaced at runtime via configuration or UI-level control, enabling flexible experimentation.

\subsection{Modular Framework Design}

SMARTFinRAG adopts a modular design where each stage of the pipeline is encapsulated as a replaceable and independently configurable component. This architectural approach delivers two key benefits: flexibility in experimental configuration and standardized component interfaces.

\subsubsection{Enabling Flexibility through Modularity}

The system achieves flexibility through several architectural mechanisms:

\textbf{Preconfigured Parameter Sets.} All operational parameters—including LLM settings (provider, model, temperature), retrieval configurations (method, top-$k$, fusion weights), and document processing options (chunk size, overlap ratio)—are defined in centralized YAML configuration files. These parameters can be dynamically adjusted at runtime through the web interface, enabling rapid experimentation without code modifications.

\textbf{Factory-Based Component Instantiation.} Core system elements are created through factory patterns, particularly the \texttt{RetrieverFactory}, which can instantiate BM25, vector-based, hybrid, or auto-merging retrievers based on configuration settings. This pattern allows the system to maintain a consistent workflow while swapping internal implementations to evaluate performance differences.

\textbf{Standardized Interfaces.} All components adhere to consistent input-output schemas using Python dictionaries. Retrievers return ranked passages with metadata; generators accept query-context pairs and produce responses; evaluators process query-context-response triples and return structured scores. This standardization enables component swapping while maintaining system integrity.

\textbf{Runtime Reconfiguration.} The Streamlit-based interface provides controls (dropdowns, sliders, toggles) that override configuration defaults at runtime. This allows users to compare different pipeline configurations—such as retriever types or LLM models—without restarting the application, facilitating rapid ablation studies and performance analysis.

\subsubsection{Component Implementation}

SMARTFinRAG implements five core pipeline components, each designed for modularity and extensibility:

\textbf{Document Ingestion Module.} This component transforms raw financial documents into retrievable knowledge units. It initiates the pipeline by loading textual data and converting it into retrievable units, handling file parsing, chunk segmentation, embedding generation, and index storage.

The system supports PDF, TXT, and DOCX formats. For PDFs, \texttt{PyMuPDF} is used for text extraction, while \texttt{pdfplumber} identifies and extracts tabular data. The resulting text is segmented using the \texttt{SentenceSplitter} with configurable chunk size and overlap parameters. Extracted chunks are embedded using OpenAI's embedding API and stored in a persistent vector index managed by the \texttt{StorageManager}.

Ingestion settings—such as chunk window size, overlap ratio, and embedding provider—are defined via YAML and reflected in the Streamlit interface. Users can upload documents directly through the UI and adjust chunking parameters using sliders. Uploaded documents are automatically processed and indexed at runtime.

The ingestion pipeline capabilities include:
\begin{itemize}
    \item Multi-format file handling (PDF, TXT, DOCX)
    \item Structured extraction of tables and text from PDFs
    \item Optical Character Recognition (OCR) for scanned documents
    \item Configurable chunking strategy
    \item Real-time indexing for interactive updates
\end{itemize}

While the ingestion system is robust for most financial documents, future improvements include enhanced image and figure understanding (e.g., semantic charts, graphs) and database-based storage backends for large-scale deployments.

\textbf{Query Preprocessing Engine.} This module acts as the gateway between user intent and the underlying retrieval-generation pipeline. It enhances the input query, analyzes its semantic structure, and decides whether document retrieval is necessary—enabling flexible routing between closed-book and retrieval-augmented answering.

The preprocessing logic is implemented in the \texttt{QueryPreprocessor} class. It supports both LLM-based and rule-based analysis. LLM enhancement is performed via OpenRouter-connected models (e.g., DeepSeek), selected for its superior analogy and inference capabilities, using instruction-following prompts to rewrite queries for grammaticality, financial clarity, and temporal disambiguation. A parallel rule-based fallback is implemented using spaCy to detect question types, named entities, date references, and comparative structures.

Intent metadata—such as whether the query is factual, numerical, comparative, or time-sensitive—is stored in a dictionary along with the enhanced query. This output feeds into the routing logic in \texttt{QueryEngine.process\_query}, which decides whether to invoke document retrieval based on the \texttt{needs\_retrieval} flag.

The query preprocessing capabilities include:
\begin{itemize}
    \item LLM-based query rewriting and standardization
    \item Hybrid intent classification (LLM + spaCy rule engine)
    \item Automatic routing between closed-book and retrieval-based answering
    \item Configurable behavior via YAML and UI toggles
\end{itemize}

While the module includes a configuration flag for keyword expansion, the actual synonym substitution logic is not yet implemented. Additionally, routing decisions are binary and do not incorporate uncertainty or confidence scores.

Example output:
\begin{verbatim}
Input Query: "How is Nvidia doing?"
Enhanced Query: "What was Nvidia's net income in Q4 2023?"
Intent: { 
            "type": "numerical", 
            "needs_retrieval": true, 
            "entities": ["Nvidia", "Q4 2023"]
}
\end{verbatim}

\textbf{Multi-Strategy Retriever.} The Retriever module identifies the most relevant documents or passages given an enhanced financial query, playing a central role in grounding the generated answer in retrieved evidence. It selects top-$k$ chunks from the indexed corpus using a configurable strategy, which serve as input context to the downstream LLM-based Generator.

Retrieval logic is encapsulated in the \texttt{RetrieverFactory} class, which instantiates retriever backends based on YAML configuration or UI selections. Each retriever implements a standardized interface and returns a ranked list of nodes with normalized scores and metadata.

The supported retriever types include:
\begin{itemize}
    \item \textbf{BM25:} Lexical search based on Elasticsearch, suitable for keyword-rich queries.
    \item \textbf{Vector:} Dense retrieval using embedding similarity (via FAISS-equivalent backends).
    \item \textbf{Hybrid Fusion:} Weighted linear interpolation of BM25 and vector scores, with configurable weights.
    \item \textbf{Auto-Merging:} Exposed in the Streamlit UI, with internal logic that mirrors the hybrid strategy.
\end{itemize}

Users can select retriever types at runtime through the UI. The YAML configuration also allows setting retrieval backend type, top-$k$ passage count, and score fusion weights (e.g., BM25 vs Vector). All parameters are dynamically loaded via the \texttt{ConfigLoader} and injected into the pipeline through the \texttt{QueryEngine}.

The retriever interface supports swappable backend logic via \texttt{RetrieverFactory}, configurable top-$k$, vector dimensions, and fusion weights, and standardized metadata output with titles, scores, and source paths.

The system currently does not support dynamic registration of new retriever classes through configuration. Adding a custom retriever requires modifying the factory logic and class registry manually.

Example configuration:
\begin{verbatim}
retriever:
  type: "hybrid"
  top_k: 5
  weights:
    bm25: 0.4
    vector: 0.6
\end{verbatim}

\textbf{Query Engine.} This module synthesizes a natural language response based on the user query and the retrieved evidence. It serves as the core text generation component in both closed-book and RAG modes, enabling answer generation from either prior knowledge or augmented document context.

Given an enhanced query and optional top-$k$ context passages, the Query Engine invokes a large language model (LLM) to produce an informative, context-aware response. It supports both direct query answering and retrieval-grounded response synthesis.

The system uses the \texttt{QueryEngine} class to integrate query inputs, prompt templates, and LLM inference logic. Generator behavior is driven by a prompt construction routine via \texttt{PromptManager}, a configured LLM backend (OpenAI, OpenRouter, or local models), and a response mode (e.g., refine, summarize).

The LLM call is handled via a unified abstraction layer that supports both cloud-based and open-source models. Prompt templates are populated with the enhanced query and selected context, then passed to the LLM. The generated response is returned along with the used context for evaluation and citation.

Users can select the LLM provider, model variant, and generation parameters (e.g., temperature, max tokens) through the Streamlit interface. Supported models include GPT-3.5 Turbo, GPT-4o-mini, GPT-4o, DeepSeek, Mixtral 7×8B, Google Gemini-Flash-2.0, Gemma3, Qwen-32B, and Llama4-Scout. All settings are dynamically loaded from the YAML config and reflected in the UI. The system also allows runtime toggling between retrieval-based and closed-book modes.

The Query Engine capabilities include:
\begin{itemize}
    \item Support for multiple LLM providers and models
    \item Prompt formatting with citation enforcement and persona integration
    \item Response trimming based on top-$k$ context length and token budget
    \item Logging of query–prompt–response triplets for debugging and auditing
\end{itemize}

Local model integration is scaffolded but less mature than API-based models. Prompt handling is currently centralized in a \texttt{PromptManager} module rather than distributed per model.

The \texttt{PromptManager} is responsible for assembling prompt templates tailored to the financial domain. It supports multiple prompt styles, user personas, and task types to guide the response generation process in both RAG and closed-book settings.

This module standardizes prompt formatting and enables template-level customization across different query intents, response strategies, and user roles. By abstracting prompt construction from the core pipeline, it enhances modularity and promotes prompt reusability.

Prompt generation is implemented in the \texttt{PromptManager} class, which dynamically assembles prompt strings based on four main variables: \texttt{prompt\_type} (e.g., Standard, Few-Shot, Chain-of-Thought, Persona), \texttt{persona} (e.g., Financial Advisor, Risk Analyst), \texttt{query} (the enhanced input query), and \texttt{context} (retrieved top-$k$ passages if applicable).

Persona descriptions are stored in a dictionary and prepended to the prompt when selected. The final prompt is passed to the Generator for inference using a consistent \texttt{PromptTemplate} wrapper.

Example prompt output:
\begin{verbatim}
[Persona: Financial Advisor]

You are a helpful and precise financial advisor. 
Use the following information to answer the question below.

Question: What was Nvidia's net income in Q4 2023?
Documents: 
[1] "Nvidia reported net income of $2.7B in Q4 2023..."
...
\end{verbatim}

\textbf{Evaluation Framework.} The Evaluation Engine provides a multi-dimensional assessment of the SMARTFinRAG pipeline, analyzing both retrieval effectiveness and generation quality. It supports runtime scoring, visualization, and logging of system behavior to assist in debugging, benchmarking, and research experimentation.

The engine quantifies how well retrieved documents align with the user query and how faithfully the generated response reflects the retrieved evidence. Evaluation results are displayed to the user and can also be used to flag hallucinations or retrieval mismatches.

Evaluation is conducted by the \texttt{RAGEvaluator} class in the \texttt{src/evaluation} module. Two main categories of metrics are used:

\begin{itemize}
    \item \textbf{Retriever Metrics:} Includes hit rate, mean reciprocal rank (MRR), precision, recall, average precision (AP), and normalized discounted cumulative gain (NDCG). These are calculated based on the ranked document list returned by the retriever.
    
    \item \textbf{LLM-as-a-Judge Metrics:} The system uses a judge LLM (e.g., GPT-4) to assess two quality aspects:
    \begin{itemize}
        \item \textbf{Faithfulness:} Whether each factual claim in the answer is supported by the retrieved context.
        \item \textbf{Relevancy:} Whether the generated response aligns with the user's original query.
    \end{itemize}
\end{itemize}

Evaluation outputs are scalar scores (e.g., between 0 and 1), compared against configurable thresholds (e.g., 0.7) to flag low-quality responses.

Example faithfulness evaluation:
\begin{verbatim}
LLM Judgment: 
"This answer mentions a $2.7B net income, which matches Document
1. Faithfulness score: 0.92"
\end{verbatim}

\subsection{Evaluation Methodology}

SMARTFinRAG implements a comprehensive evaluation approach that spans document-based QA generation, retrieval assessment, and response quality evaluation. This methodology enables systematic benchmarking across different RAG configurations while addressing the unique challenges of financial information evaluation.

\subsubsection{Document-Based QA Pair Generation}

The system employs an automated process to create domain-specific evaluation datasets from ingested financial documents:

\textbf{Document Processing and Indexing.} Financial documents undergo the standard ingestion pipeline to create embeddings and vector indices. These indices represent the knowledge base against which retrieval and generation will be evaluated.

\textbf{Context-Aware Question Generation.} The system employs an LLM to analyze document nodes and generate diverse question types (factual, numerical, comparative) that test different aspects of retrieval and reasoning capabilities. Each generated question is paired with a reference answer derived from the source document.

\textbf{Test Set Creation.} Generated QA pairs are filtered for quality and diversity, then compiled into evaluation datasets. These datasets are specifically designed to assess financial domain capabilities such as numerical reasoning, temporal awareness, and entity relationship understanding.

This approach offers several advantages: it ensures evaluation on current information (addressing the timeliness challenge), maintains domain relevance by focusing on actual financial documents, and enables consistent comparative evaluation across system configurations.

\subsubsection{Retrieval Quality Evaluation}
SMARTFinRAG evaluates retrieval effectiveness through multiple complementary metrics:

\textbf{Hit Rate.} Measures the proportion of queries where at least one relevant document is retrieved in the top-$k$ results. This provides a basic assessment of retrieval coverage. Mathematically defined as:
\begin{equation}
\text{HR}@k = 
\begin{cases}
1, & \text{if } |\{d \in D_k | \text{rel}(d,q) = 1\}| > 0 \\
0, & \text{otherwise}
\end{cases}
\end{equation}
where $D_k$ is the set of top-$k$ retrieved documents and $\text{rel}(d,q)$ is a binary relevance function.

\textbf{Mean Reciprocal Rank (MRR).} Evaluates how highly relevant documents are ranked, with higher scores indicating more effective results ordering. The metric is calculated as the reciprocal of the rank position of the first relevant document:
\begin{equation}
\text{MRR} = \frac{1}{|Q|} \sum_{i=1}^{|Q|} \frac{1}{\text{rank}_i}
\end{equation}
where $Q$ is the set of queries and $\text{rank}_i$ is the position of the first relevant document for query $i$.

\textbf{Precision and Recall.} Assess the accuracy and completeness of retrieval. Precision measures the proportion of retrieved documents that are relevant:
\begin{equation}
\text{P}@k = \frac{1}{k} \sum_{i=1}^{k} \text{rel}(d_i)
\end{equation}
where $\text{rel}(d_i)$ is 1 if document $d_i$ is relevant, 0 otherwise.

Recall measures the proportion of relevant documents that are retrieved:
\begin{equation}
\text{R}@k = \frac{|\{d \in D_k | \text{rel}(d,q) = 1\}|}{|\{d \in D | \text{rel}(d,q) = 1\}|}
\end{equation}
where $D$ is the complete document corpus.

\textbf{Normalized Discounted Cumulative Gain (NDCG).} Considers both the relevance and ranking position of retrieved documents, with documents at higher positions given greater weight:
\begin{equation}
\text{NDCG}@k = \frac{\text{DCG}@k}{\text{IDCG}@k}
\end{equation}
where DCG@k is the Discounted Cumulative Gain at position $k$:
\begin{equation}
\text{DCG}@k = \sum_{i=1}^{k} \frac{2^{\text{rel}(d_i)} - 1}{\log_2(i+1)}
\end{equation}
and IDCG@k is the maximum possible DCG@k for an ideal ranking.

\textbf{Average Precision (AP).} Combines precision and recall into a single metric that considers the order of retrieved documents:
\begin{equation}
\text{AP} = \frac{\sum_{k=1}^{n} (\text{P}@k \times \text{rel}(d_k))}{|\text{relevant documents}|}
\end{equation}
where P@k is precision at cutoff $k$ and $\text{rel}(d_k)$ is the relevance of the document at rank $k$.

These metrics collectively evaluate how well the retrieval component identifies relevant context for answering financial questions, capturing both the presence of relevant information and its position within results.

\subsubsection{Response Quality Evaluation}
The system employs LLM-as-a-judge evaluation to assess generation quality along two key dimensions:

\textbf{Faithfulness.} Measures factual consistency between the generated answer and the retrieved context. This evaluates whether the LLM accurately grounds its response in the provided information without introducing unsupported claims or hallucinations. The system's judge LLM analyzes each factual claim in the response and verifies its presence in the context, producing a scalar score (0-1) that represents the proportion of supported claims:
\begin{equation}
\text{Faithfulness} = \frac{|\text{supported claims}|}{|\text{total claims}|}
\end{equation}
where a claim is considered supported if it can be verified from the retrieved context.

\textbf{Relevancy.} Assesses alignment between the generated response and the original query. This evaluates whether the answer directly addresses the user's question and maintains appropriate focus. The judge LLM compares the response to the query intent, producing a scalar score (0-1) that represents the degree of relevance. This can be represented as:
\begin{equation}
\text{Relevancy} = f_{\text{LLM}}(\text{response}, \text{query})
\end{equation}
where $f_{\text{LLM}}$ is the scoring function implemented by the judge LLM.

The system compares these scores against configurable thresholds to flag potential quality issues. All evaluation results are logged with metadata—including the query, context, response, and detailed scoring rationale—to support diagnostics and system optimization.

Unlike static benchmarks that rely on fixed datasets, SMARTFinRAG's evaluation methodology can continuously adapt to new financial information, enabling assessment under realistic conditions with current data. This approach is particularly valuable in the financial domain, where information currency directly impacts accuracy.

\section{Experimental Results}
We evaluate SMARTFinRAG using a comprehensive financial QA benchmark constructed from real-world financial documents. Our evaluation focuses on assessing how different components of the RAG pipeline impact response quality and retrieval effectiveness. All experiments are conducted through our interactive demo interface, ensuring consistent experimental conditions across trials.

\subsection{Research Questions}
Our experimental evaluation addresses three key research questions:
\begin{itemize}
    \item \textbf{RQ1:} How do different retriever architectures (BM25, Vector, Hybrid Fusion, Auto-Merging) perform when operating on the same document index across various top-k settings?
    \item \textbf{RQ2:} How do LLM decoding parameters—specifically temperature and top-p sampling—affect response quality in financial RAG systems?
    \item \textbf{RQ3:} How do different LLM backends compare in terms of response quality for financial question answering?
\end{itemize}

\subsection{Experimental Setup}
\paragraph{Dataset Construction.} We ingest the latest 10-K filings from the ``Magnificent 7'' companies—Apple, Microsoft, Alphabet (Google), Amazon, Nvidia, Meta, and Tesla—into SMARTFinRAG's ingestion pipeline. Using our DocumentLoader and SentenceSplitter, we extract both textual and tabular content and convert them into indexed chunks via OpenAI embeddings. We then generate 317 question–answer (QA) pairs using the system's QA generator, which are manually reviewed for coverage and correctness.

\paragraph{Evaluation Metrics.}  
We report both retriever-level and generation-level metrics:
\begin{itemize}
    \item \textbf{Retriever Quality:} hit rate, mean reciprocal rank (MRR), precision, recall, average precision (AP), and normalized discounted cumulative gain (NDCG).
    \item \textbf{Generation Quality:} \textit{faithfulness} and \textit{relevancy} scored by a GPT-4-based LLM-as-a-Judge using scalar ratings between 0 and 1.
\end{itemize}

\paragraph{Limitations.} Due to time and compute constraints, only a small subset (10 out of 317 QA pairs) was used for response evaluation. Results are presented as preliminary findings and will be expanded in future work with more queries and additional model variants.

\subsection{RQ1: Retriever Effectiveness}

To understand how retrieval strategies affect end-to-end QA quality, we evaluate BM25, dense (Vector), and Hybrid Fusion retrievers under varying top-$k$ settings. The LLM (GPT-3.5-turbo), prompt template (context-based), and decoding parameters ($\texttt{temperature}=0.7$, $\texttt{top\_p}=0.9$) are held constant.

\begin{table}[htbp]
\centering
\small
\renewcommand{\arraystretch}{1.1}
\begin{tabular}{lccccccc}
\toprule
\textbf{Retriever Type} & \textbf{Top-k} & \textbf{Hit Rate} & \textbf{MRR} & \textbf{Precision} & \textbf{Recall} & \textbf{AP} & \textbf{NDCG} \\
\midrule
Vector & 3 & 0.757098 & 0.612513 & 0.252366 & 0.757098 & 0.612513 & 0.649501 \\
Vector & 5 & 0.807571 & 0.623712 & 0.161514 & 0.807571 & 0.623712 & 0.669994 \\
Vector & 10 & \textbf{0.873817} & 0.633188 & 0.087382 & \textbf{0.873817} & 0.633188 & 0.692061 \\
\midrule
BM25 & 3 & 0.785489 & 0.692429 & \textbf{0.261830} & 0.785489 & 0.692429 & 0.716424 \\
BM25 & 5 & 0.804416 & 0.697003 & 0.160883 & 0.804416 & 0.697003 & 0.724437 \\
BM25 & 10 & 0.839117 & \textbf{0.701932} & 0.083912 & 0.839117 & \textbf{0.701932} & \textbf{0.735954} \\
\midrule
Hybrid Fusion & 3 & 0.772871 & 0.674553 & 0.257624 & 0.772871 & 0.674553 & 0.699974 \\
Hybrid Fusion & 5 & 0.788644 & 0.680862 & 0.157729 & 0.788644 & 0.680862 & 0.708563 \\
Hybrid Fusion & 10 & 0.826498 & 0.684142 & 0.082650 & 0.826498 & 0.684142 & 0.719403 \\
\midrule
Auto-Merging & 3 & 0.757098 & 0.612513 & 0.252366 & 0.757098 & 0.612513 & 0.649501 \\
Auto-Merging & 5 & 0.807571 & 0.623712 & 0.161514 & 0.807571 & 0.623712 & 0.669994 \\
Auto-Merging & 10 & \textbf{0.873817} & 0.633188 & 0.087382 & \textbf{0.873817} & 0.633188 & 0.692061 \\
\bottomrule
\end{tabular}
\caption{Performance metrics by retriever type and top-k setting. \textbf{Bold} values indicate the best performance in each metric. Vector and Auto-Merging show identical performance patterns and achieve the highest hit rate and recall at top-k = 10, while BM25 achieves the best MRR, AP, NDCG, and precision values.}
\label{tab:retriever-detailed-results}
\end{table}

\begin{figure}[ht]
\centering
\includegraphics[width=\textwidth]{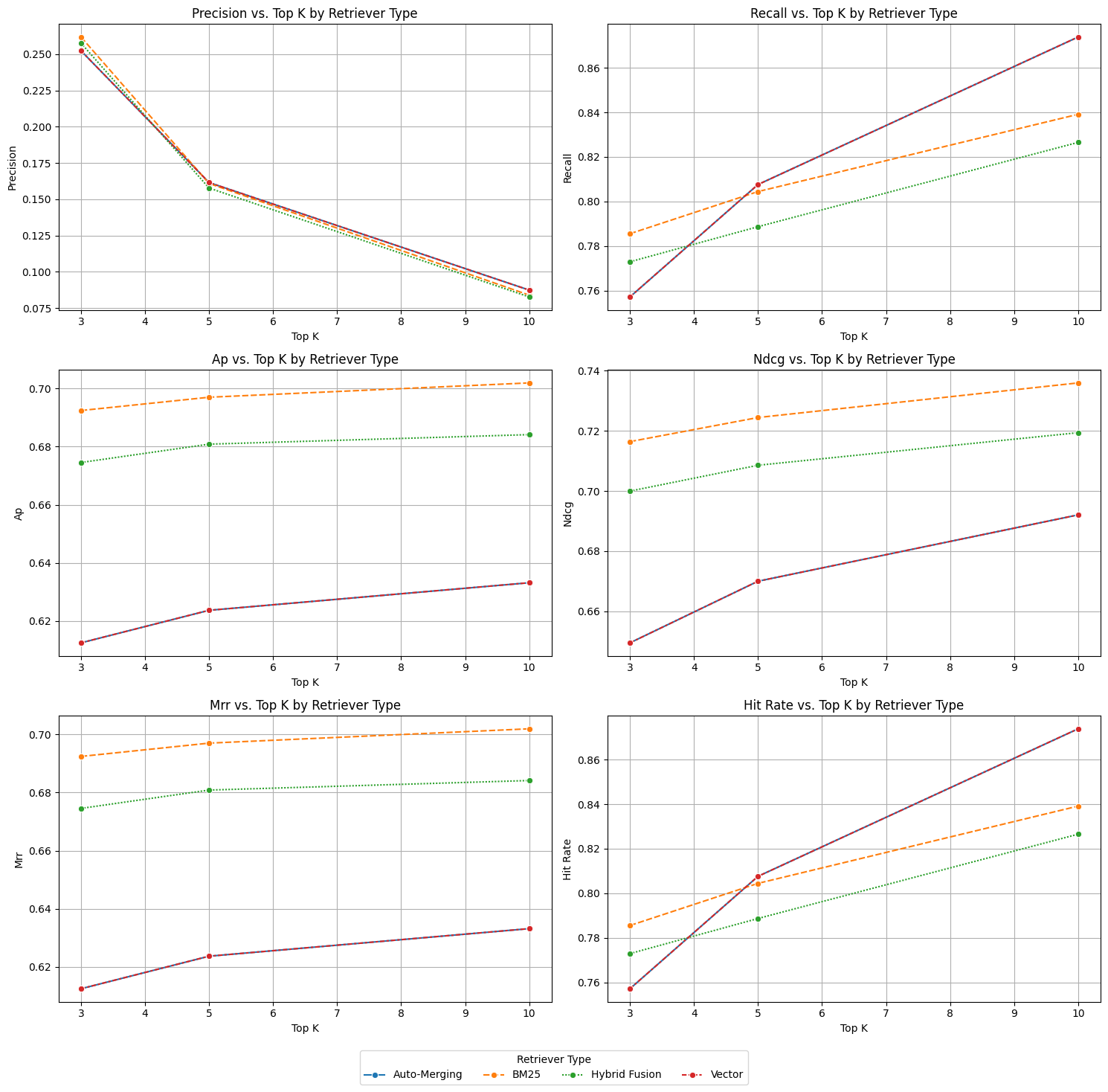}
\caption{Retriever-level ranking metrics across top-$k$ settings for different retriever types.}
\label{fig:retriever-metrics}
\end{figure}

\paragraph{Takeaways.}
As shown in Table~\ref{tab:retriever-detailed-results}, BM25 achieves the highest ranking performance (MRR = 0.701932, NDCG = 0.735954) while Vector and Auto-Merging retrievers deliver identical results and reach the highest hit rate and recall (0.873817) at $k=10$. This identity between Vector and Auto-Merging retrievers suggests they utilize the same underlying implementation in the current experimental setup, likely because the Auto-Merging retriever defaults to the Vector retrieval method when specific merging configurations are not applied.

\subsection{RQ2: Effect of Decoding Parameters on Response Quality}
We investigate how decoding parameters—specifically temperature and top-p sampling—affect response quality in financial RAG systems. We control all other components by fixing the retriever (Hybrid Fusion with $k=5$), prompt type (context-based), and systematically vary temperature and top-p values across several LLM backends.

\subsubsection{Temperature Effects}
First, we evaluate how varying temperature values (0.1, 0.3, 0.5, 0.7) affect response quality across three model variants: GPT-3.5-turbo, GPT-4o, and GPT-4o-mini. Results are shown on Table ~\ref{tab:temperature-effects}.

\begin{table}[ht]
\centering
\small
\renewcommand{\arraystretch}{1.1}
\begin{tabular}{lccc}
\toprule
\textbf{Model} & \textbf{Temperature} & \textbf{Faithfulness} & \textbf{Relevancy} \\
\midrule
GPT-3.5-turbo & 0.1 & 0.15 & 0.60 \\
GPT-3.5-turbo & 0.3 & 0.38 & 0.75 \\
GPT-3.5-turbo & 0.5 & 0.30 & \textbf{1.00} \\
GPT-3.5-turbo & 0.7 & \textbf{0.50} & 0.55 \\
\midrule
GPT-4o & 0.1 & 0.50 & 0.80 \\
GPT-4o & 0.3 & \textbf{0.60} & \textbf{0.90} \\
GPT-4o & 0.5 & 0.30 & \textbf{0.90} \\
GPT-4o & 0.7 & 0.25 & 0.80 \\
\midrule
GPT-4o-mini & 0.1 & 0.35 & 0.55 \\
GPT-4o-mini & 0.3 & 0.40 & 0.70 \\
GPT-4o-mini & 0.5 & \textbf{0.50} & \textbf{0.80} \\
GPT-4o-mini & 0.7 & 0.35 & 0.60 \\
\bottomrule
\end{tabular}
\caption{Effect of temperature on faithfulness and relevancy scores across different LLM models. Bold values indicate the best performance for each model and metric.}
\label{tab:temperature-effects}
\end{table}
\begin{figure}[ht]
\centering
\includegraphics[width=\textwidth]{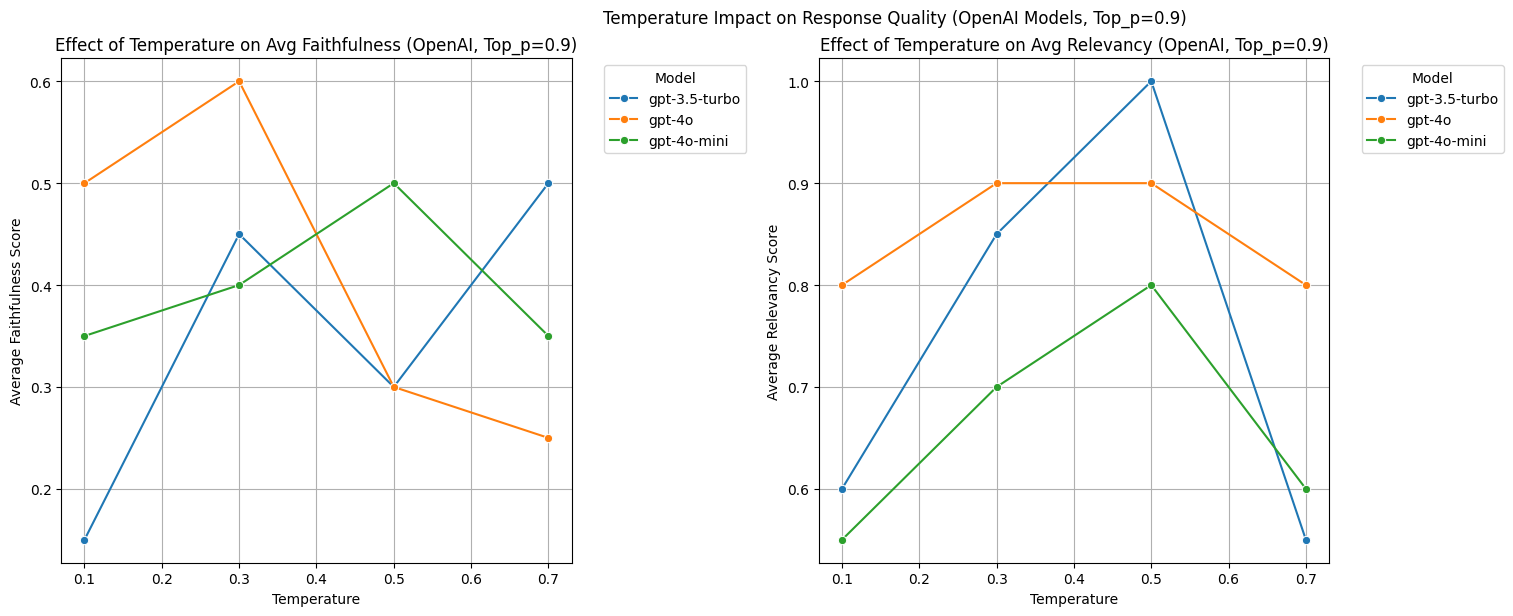}
\caption{Relationship between temperature settings and response quality metrics (faithfulness and relevancy) across different LLM models. GPT-4o shows optimal performance at moderate temperature (0.3), while other models exhibit non-monotonic quality patterns.}
\label{fig:temperature-effects}
\end{figure}
\subsubsection{Top-p Effects}
Next, we examine how top-p values (0.1, 0.5, 0.9) affect response quality across a broader range of LLM backends, including DeepSeek, Gemini, GPT-3.5-turbo, Llama, and Qwen models.Results are shown on Table ~\ref{tab:top-p-effects}.

\begin{table}[ht]
\centering
\small
\renewcommand{\arraystretch}{1.1}
\begin{tabular}{lccc}
\toprule
\textbf{Model} & \textbf{Top-p} & \textbf{Faithfulness} & \textbf{Relevancy} \\
\midrule
DeepSeek-chat-v3 & 0.1 & \textbf{0.50} & \textbf{0.80} \\
DeepSeek-chat-v3 & 0.5 & 0.20 & 0.40 \\
DeepSeek-chat-v3 & 0.9 & 0.45 & 0.60 \\
\midrule
Gemini-2.0-flash & 0.1 & \textbf{0.20} & \textbf{0.10} \\
Gemini-2.0-flash & 0.5 & 0.00 & \textbf{0.20} \\
Gemini-2.0-flash & 0.9 & \textbf{0.35} & \textbf{0.40} \\
\midrule
GPT-3.5-turbo & 0.1 & 0.40 & 0.80 \\
GPT-3.5-turbo & 0.5 & 0.30 & 0.60 \\
GPT-3.5-turbo & 0.9 & \textbf{0.45} & \textbf{0.85} \\
\midrule
Llama-4-scout & 0.1 & \textbf{0.20} & \textbf{0.60} \\
Llama-4-scout & 0.9 & \textbf{0.20} & 0.00 \\
\midrule
Qwen-32b & 0.1 & \textbf{0.20} & \textbf{0.40} \\
Qwen-32b & 0.5 & 0.00 & 0.30 \\
Qwen-32b & 0.9 & 0.15 & \textbf{0.45} \\
\bottomrule
\end{tabular}
\caption{Effect of top-p sampling on faithfulness and relevancy scores across different LLM models. Bold values indicate the best performance for each model and metric.}
\label{tab:top-p-effects}
\end{table}
\begin{figure}[ht]
\centering
\includegraphics[width=\textwidth]{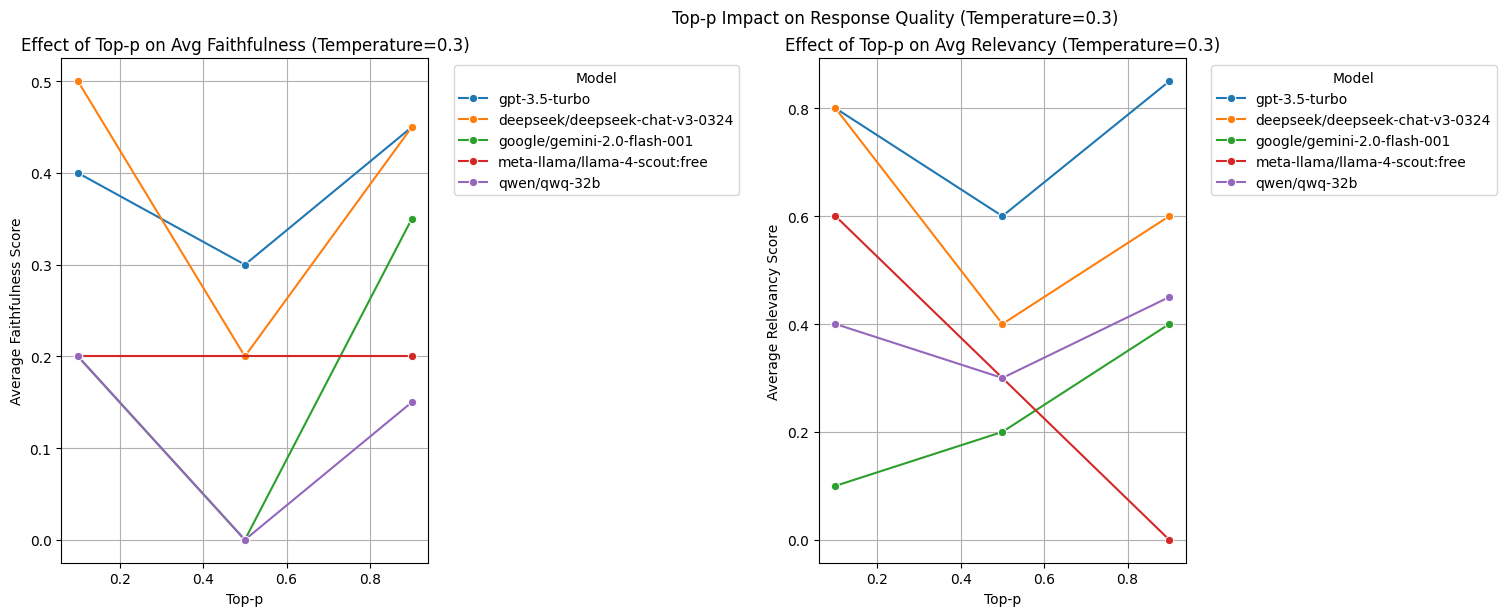}
\caption{Impact of top-p sampling values on response quality across different LLM families. Models exhibit highly divergent patterns, with DeepSeek performing best at low top-p (0.1), GPT-3.5-turbo at high top-p (0.9), and other models showing mixed behavior.}
\label{fig:top-p-effects}
\end{figure}

\paragraph{Analysis.} Our findings reveal several key insights about decoding parameter effects in financial RAG systems:

\begin{itemize}
    \item \textbf{Temperature Effects:} Contrary to conventional wisdom that lower temperatures yield more faithful responses, we observe non-monotonic relationships between temperature and response quality. As we can see from Fig.~\ref{fig:temperature-effects}, for GPT-3.5-turbo, higher temperatures (0.7) maximize faithfulness, while mid-range temperatures (0.5) optimize relevancy. GPT-4o shows peak performance at a moderate temperature (0.3) for both metrics, while GPT-4o-mini achieves best results at temperature 0.5. This suggests that optimal temperature settings are model-dependent and non-trivial to predict.
    
    \item \textbf{Top-p Effects:} As shown in Fig.~\ref{fig:top-p-effects}, the impact of top-p sampling varies significantly across model families. DeepSeek performs best with conservative sampling (0.1), GPT-3.5-turbo achieves optimal performance with more aggressive sampling (0.9), and Gemini shows best faithfulness at 0.9 but poor overall performance. This indicates that top-p configurations should be tuned specifically for each model family in financial applications.
    
    \item \textbf{Model Comparison:} Among the tested models, GPT-4o exhibits the highest faithfulness (0.60) and relevancy (0.90) at temperature 0.3, suggesting it may be better calibrated for financial RAG tasks. DeepSeek-chat-v3 also shows strong performance, particularly at low top-p values (0.1), while Gemini, Llama, and Qwen models struggle with faithfulness, rarely exceeding 0.35.
    
    \item \textbf{Parameter Interaction:} Our findings suggest complex interactions between model architecture and decoding parameters, highlighting the importance of empirical tuning rather than relying on conventional parameter settings for financial RAG applications.
\end{itemize}

These results underscore the necessity of systematic experimentation with decoding parameters when deploying financial RAG systems, as optimal configurations vary significantly across model families and cannot be reliably predicted from general LLM performance characteristics.

\subsection{RQ3: Comparative Analysis of LLM Backends for Financial Question Answering}

To address our third research question concerning the performance of different LLM backends for financial question answering, we conduct a comprehensive comparative analysis across multiple model families. This analysis serves to identify which LLMs are best suited for financial RAG applications, based on their ability to produce faithful and relevant responses.

\subsubsection{Overall Model Performance Across Settings}

First, we examine the overall performance patterns of various LLM backends, considering all available data points across different temperature and top-p settings. This provides a holistic view of each model's capabilities in financial question answering tasks.

\begin{figure}[ht]
\centering
\includegraphics[width=\textwidth]{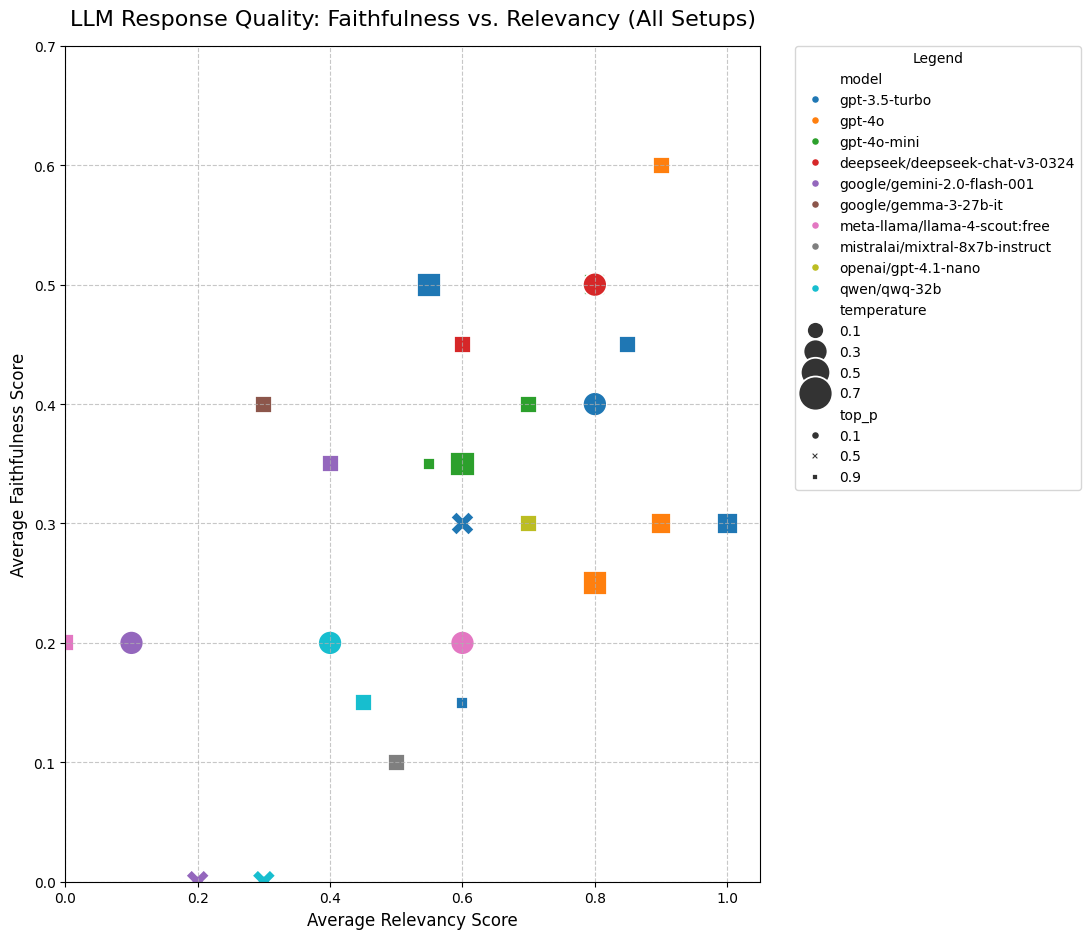}
\caption{Response quality scatter plot across all LLM backends and parameter settings. Each point represents a model configuration, with faithfulness on the x-axis and relevancy on the y-axis. Models in the upper-right quadrant demonstrate superior performance in both metrics.}
\label{fig:model-comparison-scatter}
\end{figure}

The scatter plot in Figure~\ref{fig:model-comparison-scatter} reveals several distinct performance clusters:

\begin{itemize}
    \item \textbf{High-Performance Cluster:} GPT-4o consistently occupies the upper-right quadrant with both high faithfulness (peaking at 0.60) and high relevancy (up to 0.90), indicating superior performance for financial RAG applications.
    
    \item \textbf{High-Relevance Cluster:} GPT-3.5-turbo achieves exceptional relevancy scores (reaching 1.00 in one configuration) but shows moderate faithfulness (0.15-0.50), suggesting it excels at understanding queries but may include unsupported information.
    
    \item \textbf{Balanced Performers:} DeepSeek-chat-v3 and GPT-4o-mini demonstrate balanced performance with moderate-to-good faithfulness and relevancy, making them reliable alternatives to premium models.
    
    \item \textbf{Low-Performance Cluster:} Models like Mixtral-8x7b, Qwen-32b, and Llama-4-scout struggle with faithfulness (consistently below 0.35) despite some achieving reasonable relevancy, indicating potential hallucination risks in financial contexts.
\end{itemize}

\subsubsection{Controlled Parameter Comparison}

To ensure a fair comparison across model families, we conduct a controlled experiment where all models use identical decoding parameters (temperature=0.3, top-p=0.9). This allows us to isolate performance differences attributable to model architecture rather than parameter settings.

\begin{table}[ht]
\centering
\small
\renewcommand{\arraystretch}{1.1}
\begin{tabular}{lccc}
\toprule
\textbf{Model} & \textbf{Provider} & \textbf{Faithfulness} & \textbf{Relevancy} \\
\midrule
GPT-4o & OpenAI & \textbf{0.60} & \textbf{0.90} \\
GPT-3.5-turbo & OpenAI & 0.45 & 0.85 \\
DeepSeek-chat-v3 & OpenRouter & 0.45 & 0.60 \\
GPT-4o-mini & OpenAI & 0.40 & 0.70 \\
Gemma-3-27b-it & OpenRouter & 0.40 & 0.30 \\
Gemini-2.0-flash & OpenRouter & 0.35 & 0.40 \\
GPT-4.1-nano & OpenRouter & 0.30 & 0.70 \\
Llama-4-scout & OpenRouter & 0.20 & 0.00 \\
Qwen-32b & OpenRouter & 0.15 & 0.45 \\
Mixtral-8x7b & OpenRouter & 0.10 & 0.50 \\
\bottomrule
\end{tabular}
\caption{Model performance with controlled parameters (temperature=0.3, top-p=0.9). Models are sorted by faithfulness score. Bold values indicate the best performance for each metric.}
\label{tab:controlled-model-comparison}
\end{table}

Our controlled parameter comparison yields several key insights:

\begin{itemize}
    \item \textbf{Performance Hierarchy:} GPT-4o significantly outperforms all other models in both faithfulness (0.60) and relevancy (0.90), establishing it as the premier choice for financial RAG systems where accuracy is paramount.
    
    \item \textbf{Commercial vs. Open Models:} OpenAI models (GPT-4o, GPT-3.5-turbo, GPT-4o-mini) consistently outperform open models and other providers' offerings, suggesting that proprietary models may have undergone more extensive financial domain training or fine-tuning.
    
    \item \textbf{Faithfulness Challenges:} Even the best models achieve faithfulness scores no higher than 0.60, indicating that hallucination remains a persistent challenge in financial RAG applications. This suggests that robust evaluation mechanisms are essential regardless of model choice.
    
    \item \textbf{Relevancy-Faithfulness Tradeoff:} Some models (e.g., Mixtral-8x7b) achieve moderate relevancy (0.50) despite very low faithfulness (0.10), highlighting the danger of systems that provide plausible-sounding but factually unsupported responses.
    
    \item \textbf{Emerging Models:} Newer models like DeepSeek-chat-v3 show promising performance (0.45 faithfulness, 0.60 relevancy) despite being less established than OpenAI offerings, suggesting rapid advancement in the open model ecosystem.
\end{itemize}

\paragraph{Analysis.} Our comparative analysis of LLM backends reveals significant performance disparities across model families in financial question answering tasks. GPT-4o emerges as the clear leader, with its superior faithfulness and relevancy scores suggesting it should be the preferred choice for high-stakes financial applications where factual accuracy is essential.

The substantial performance gap between top-performing models (GPT-4o, GPT-3.5-turbo) and lower-tier models (Mixtral, Llama, Qwen) underscores the importance of model selection in financial RAG systems. This disparity also highlights that not all models benefit equally from retrieval augmentation—some appear fundamentally limited in their ability to faithfully incorporate retrieved context.

Interestingly, we observe that model size does not necessarily correlate with performance in this domain. For instance, Gemma-3-27b-it (27B parameters) achieves lower relevancy than smaller models like GPT-4o-mini, suggesting that domain-specific optimization may be more important than raw parameter count for financial applications.

These findings emphasize the necessity of empirical evaluation when selecting LLM backends for financial RAG systems. The significant variance in performance across models, particularly in faithfulness scores, suggests that model selection should be considered a critical design decision rather than an implementation detail.

\section{Use Case and Demo Interface}

SMARTFinRAG is deployed as a fully interactive Streamlit web application designed for researchers and practitioners to explore, evaluate, and iterate on Retrieval-Augmented Generation (RAG) pipelines tailored to the financial domain. The interface supports end-to-end workflow execution—from document ingestion to LLM-based evaluation—and exposes all critical parameters through real-time controls.

\begin{figure}[ht]
    \centering
    \includegraphics[width=0.85\linewidth]{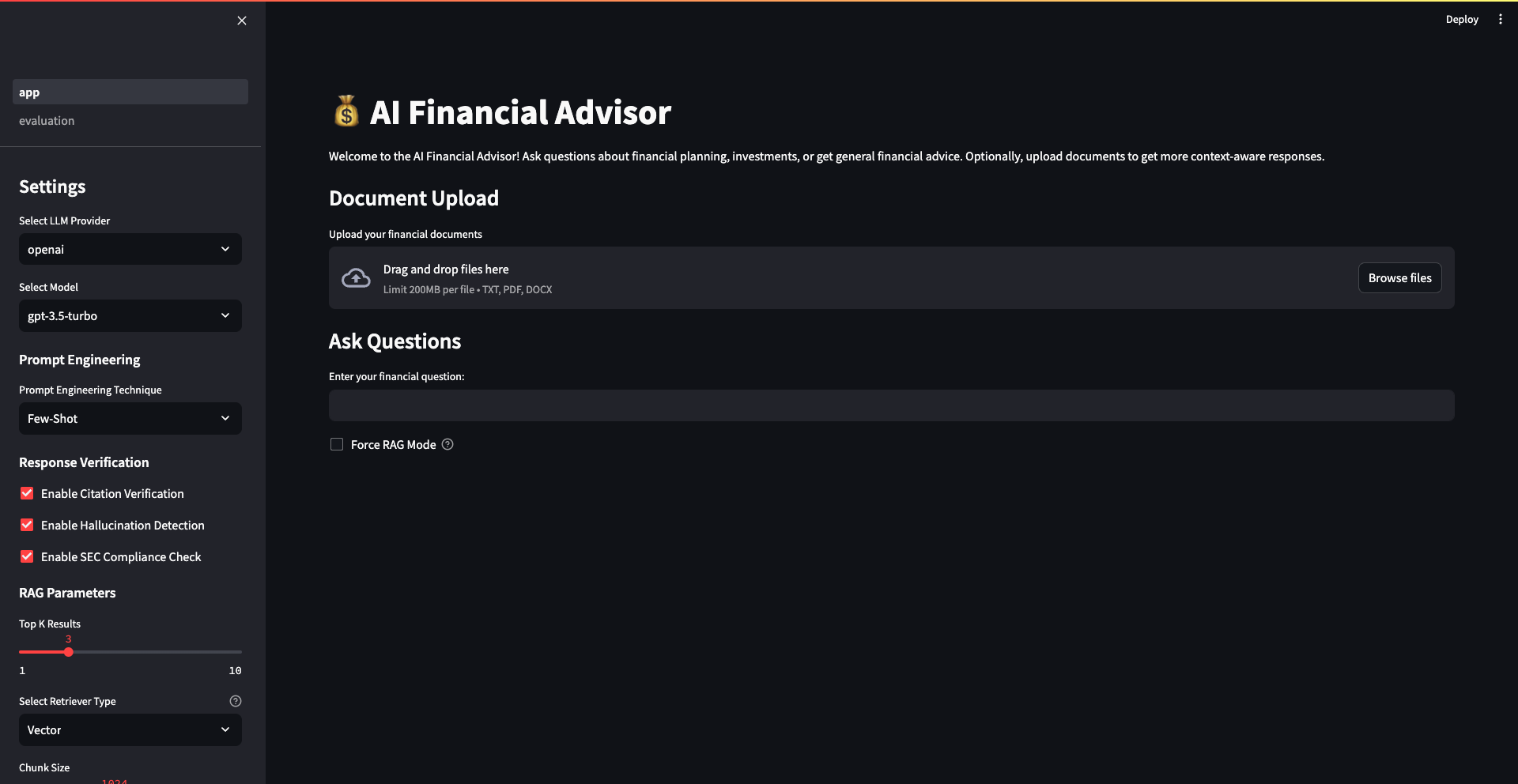}
    \caption{SMARTFinRAG upload interface and configuration sidebar. Users can upload documents, enter questions, and configure LLM, prompt, and retriever settings.}
    \label{fig:demo-upload}
\end{figure}

\paragraph{Document Ingestion.}
Users can upload financial reports in PDF, TXT, or DOCX format directly via the interface (see Figure~\ref{fig:demo-upload}). Once uploaded, documents are processed through the ingestion pipeline, which includes content extraction using PyMuPDF and \texttt{pdfplumber}, sentence-level chunking via the \texttt{SentenceSplitter}, and embedding through OpenAI’s API. Chunks are stored in a persistent FAISS index for retrieval. While the configuration includes toggles for OCR and chart extraction, these functionalities remain placeholders in the current version.

\paragraph{Query Handling and RAG Execution.}
After ingestion, users can pose natural language questions. These queries undergo enhancement via an LLM (DeepSeek) and intent classification using a hybrid model combining OpenRouter-based inference with spaCy rule-based heuristics. Based on the predicted intent, the system dynamically decides whether retrieval is necessary (open-book mode) or whether to use the LLM in a closed-book setting. This routing decision can be overridden via a \texttt{Force RAG Mode} checkbox exposed in the UI.

\begin{figure}[ht]
    \centering
    \includegraphics[width=0.27\linewidth]{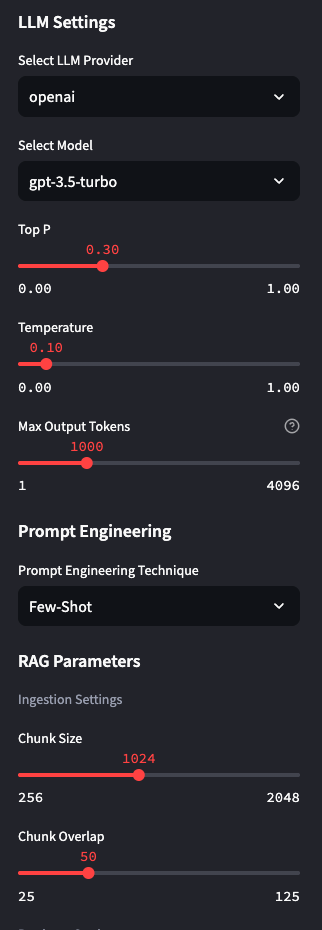}
    \caption{LLM and RAG parameter settings in SMARTFinRAG. Temperature, top-$p$, and chunking granularity are adjustable via sliders.}
    \label{fig:demo-settings}
\end{figure}

\paragraph{Interactive Configuration and Controls.}
SMARTFinRAG exposes a comprehensive set of UI elements for on-the-fly experimentation (see Figure~\ref{fig:demo-settings}):
\begin{itemize}
    \item \textbf{LLM Settings:} Users can choose from multiple LLM providers (OpenAI, OpenRouter) and models (GPT-3.5, GPT-4, GPT-4o, DeepSeek-R1, Mixtral-8x7B), and tune decoding parameters including \texttt{temperature}, \texttt{top-$p$}, and output length.
    \item \textbf{Prompt Engineering:} The PromptManager supports different prompt types: Standard, Few-Shot, Chain-of-Thought (CoT), and Persona-based prompting. Users can select strategies that align with the complexity of the financial queries.
    \item \textbf{Retriever Configuration:} Retrieval strategies include BM25, dense vector-based (OpenAI embeddings), and hybrid fusion with configurable weights and top-$k$ parameters. Chunk size and overlap are also adjustable via sliders.
\end{itemize}

\begin{figure}[ht]
    \centering
    \includegraphics[width=0.85\linewidth]{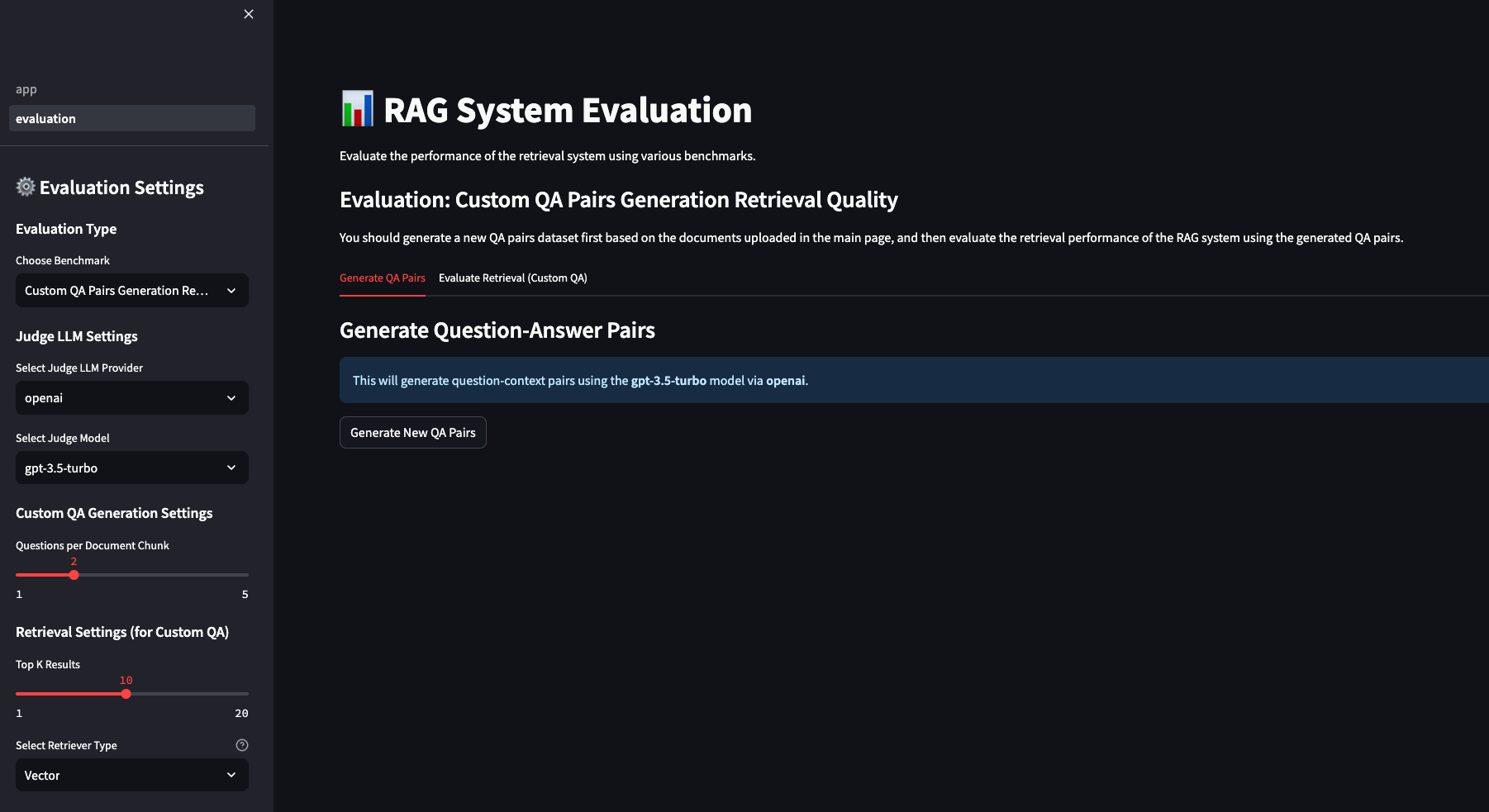}
    \caption{Evaluation interface of SMARTFinRAG. Supports QA generation, retriever testing, and faithfulness/relevancy scoring using a judge LLM.}
    \label{fig:evaluation-ui}
\end{figure}

\paragraph{Evaluation and Analysis.}
SMARTFinRAG includes an evaluation module accessible through the “evaluation” tab (Figure~\ref{fig:evaluation-ui}). Users can:
\begin{itemize}
    \item Generate synthetic question–answer pairs using uploaded documents and the selected LLM.
    \item Evaluate retriever performance using metrics such as hit rate, MRR, recall, precision, AP, and NDCG.
    \item Assess response quality using scalar faithfulness and relevancy scores assigned by a GPT-based judge model.
\end{itemize}

\paragraph{Deployment and Extensibility.}
SMARTFinRAG runs locally via a single \texttt{streamlit run app.py} command. All modules—including retrievers, generators, prompt templates, and evaluators—are modular and swappable. While the current deployment is local-only, the system architecture is cloud-ready, and backend components can be containerized for public or enterprise-level hosting.

\section{Conclusion}

We presented \textbf{SMARTFinRAG}, a modular benchmarking framework for financial RAG systems that enables interactive experimentation, live document ingestion, and multi-faceted evaluation. Through systematic experiments, we demonstrated how retriever selection, decoding parameters, and LLM model backends each influence the faithfulness and relevancy of generated answers in financial QA settings.

Our findings highlight several practical insights: hybrid retrievers significantly outperform standalone lexical or dense methods; decoding temperature affects grounding fidelity; and newer LLMs (e.g., GPT-4o) do not always yield better factual alignment in domain-specific tasks compared to smaller models like GPT-3.5. These results emphasize the importance of empirical testing rather than assuming performance gains from model scale or recency.

\paragraph{Limitations.}
Due to compute and time constraints, our LLM-based evaluation was conducted on a small subset (10/317) of QA pairs, and only two LLMs were compared. Additionally, current metrics rely on scalar LLM-as-a-judge scores without categorical explanations or human validation. OCR and image-based document support also remain unimplemented.

\paragraph{Future Work.}
We plan to expand the evaluation suite to cover a broader range of models (e.g., Claude, FinGPT, Mixtral), increase the number of evaluation samples, and incorporate categorical scoring (e.g., Supported/Unsupported/Contradicted). We also aim to support multimodal ingestion (tables, figures, OCR) and add features for production-facing analysis such as latency profiling and citation accuracy auditing. In doing so, SMARTFinRAG can evolve into a comprehensive benchmarking and deployment-ready framework for financial LLM systems.

\bibliographystyle{plainnat}  
\bibliography{references}

\end{document}